# Learnings from an Under the Hood Analysis of an Object Storage Node IO Stack

Pratik Mishra, Rekha Pitchumani, and Yang-Suk Kee, *Samsung Semiconductor Inc., USA*.

**Abstract**—Conventional object-stores are built on top of traditional OS storage stack, where I/O requests typically transfers through multiple hefty and redundant layers. The complexity of object management has grown dramatically with the ever increasing requirements of performance, consistency and fault-tolerance from storage subsystems. Simply stated, more number of intermediate layers are encountered in the I/O data path, with each passing layer adding its own syntax and semantics. Thereby increasing the overheads of request processing. In this paper, through comprehensive under-the-hood analysis of an object-storage node, we characterize the impact of object-store (and user-application) workloads on the OS I/O stack and its subsequent rippling effect on the underlying object-storage devices (OSD). We observe that the legacy architecture of the OS based I/O storage stack coupled with complex data management policies leads to a performance mismatch between what an end-storage device is capable of delivering and what it actually delivers in a production environment. Therefore, the gains derived from developing faster storage devices is often nullified. These issues get more pronounced in highly concurrent and multiplexed cloud environments. Owing to the associated issues of object-management and the vulnerabilities of the OS I/O software stacks, we discuss the potential of a new class of storage devices, known as Object-Drives. *Samsung Key-Value SSD (KV-SSD)* [1] and *Seagate Kinetic Drive* [2] are classic industrial implementations of object-drives, where host data management functionalities can be offloaded to the storage device. This leads towards the simplification of the over-all storage stack. Based on our analysis, we believe object-drives can alleviate object-stores from highly taxing overheads of data management with 20-38% time-savings over traditional *Operating Systems* (OS) stack.

**Index Terms**— Object-storage, Distributed systems, File Systems, Operating System, Linux, Storage devices, Key-Value SSD (KV-SSD), Solid State Drives (SSDs), Storage I/O Stack.

. — — — — — — — ◆ — — — — — — — .

## 1  INTRODUCTION

Object-storage, Distributed systems, File Systems, Operating System, Linux, Storage devices, Key-Value SSD (KV-SSD), Solid State Drives (SSDs), Storage I/O Stack [3]–[6] . Nearly all major cloud providers offer object-storage services for cloud remote storage with some of the most popular choices being AWS S3, Google Cloud Storage, Azure Blobstore, Ceph, Lustre, MinIO, Swift, Alibaba Cloud OSS, IBM Cloud, etc.c[3], [4], [7]–[9].

A highly simplistic representation of the software stacks in a conventional object-store is shown (a). *Objects* and their associated metadata is accessible to applications

---

- *Pratik Mishra, Rekha Pitchumani, and Yang-Suk Kee. Samsung Semiconductor, Inc. San Jose, CA.*
  *E-mail: {mishra.p, r.pitchumani, yangseok.ki@samsung.com}.*

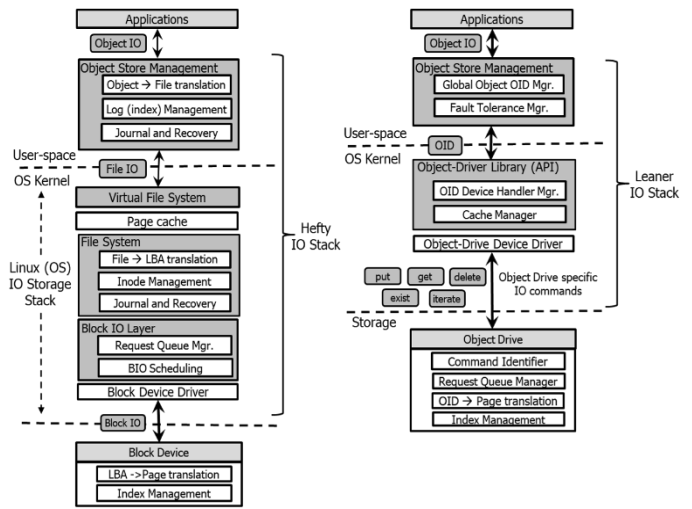

(a) Conventional (OS) Object-Store       (b) Object-Drive Object-Store

Figure 1 Object-storage I/O stack based on (a) Conventional (OS) block devices; (b) Object-Drive. More the intermediate layers in the I/O path, higher is the delay.

via Object-Storage (REST) API. Object-stores receive application requests over network, which are processed by Object-Store Management services (OSMs). OSMs provide an interface between applications and storage, thereby responsible for data management. However the actual IO is processed by backend storage servers which have locally attached storage devices known as Object-Storage Devices (OSDs) for persisting objects. Storage nodes are built on top of legacy operating system (OS) stacks due to the maturity and convenience in production environments. The OS acts as a middleware between distributed OSM layer and the OSDs. Therefore, *the performance of the object-store at the distributed level is significantly impacted by the local node operating system.*

Consider *Figure 1*(a), we also observe that along the odyssey of data access i.e. from user-applications to OSDs, IO requests transfers through multiple hefty and redundant layers. Each layer has its own syntax and semantics which consumes host resources such as compute, memory, caches, etc. for both data and metadata. The compounding effect of all the intermediate layers and their dependencies shapes the final IO workloads observed at the OSD interface. For e.g.: multiple address translations are required to serve an IO request, i.e. Object-IO → File-IO →Block-IO→Physical Page Address (PPA). Simply stated, *more the intermediate layers encountered in the IO data path, higher are the delays in request processing.*

In this paper, through comprehensive theoretical and empirical analysis we characterize the impact of complex data-management policies and user-application workloads on an in-node traditional OS backed object-storage server (see Section 2). Particularly in the OS- the Virtual File System (VFS), File-System, and the Block IO layer are the most critical components which enable data access and persistence. It is the orchestration and design of these layers which influence the over-all performance of the storage system, i.e. how efficiently objects are managed. Though object stores were initially designed for Hard Disk Drives (HDDs) but with changing customer requirements most cloud providers also offer Solid State Drive (SSD) backends as well [9], [10] . Our goal is to demonstrate and quantify the impact of the OS overheads in object-stores, more specifically *what, why, and, how* leads to the underutilization of the OSDs deployed, i.e. HDDs and SSDs both. We observe that the gains derived from storage devices is typically nullified in highly concurrent, and multiplexed cloud object-store environments due to the design principles of traditional OS stacks. Therefore, *object-store suffers from sub-optimal performance and high host resource footprint for its entire life-time.*

In Section 3, we identify the major challenges and bottlenecks in object-storage servers, such as (1) multilayer translations and index management; (2) resource contentions. Through our findings, we observe that the afore-mentioned issues leads to *fragmentation or aging[1]* of the physical file-system. This leads towards sub-optimal data placement, i.e. non-contiguous address space or Logical Block Addresses (LBA), for current and future data accesses and persistence. For the host (and user-applications) this effect translates into higher resource consumption and delays due to more translations, cache and disk activity, and book-keeping overheads.

Considering the complexity of building storage systems from scratch, a natural progression is the development of holistic storage device technologies which can execute host data management functionalities. We discuss the potential of such a class of storage devices, defined as **Object-Drives** as solution to alleviate object-stores from the associated issues of the redundant and bulky IO stack. Samsung KV-SSD [1] and Seagate Kinetic Drive [2] are classic industrial use-cases implementations of Object-Drives. Enabling *object-drives* in data-centric environments can simplify and streamline the overall data management software stack while balancing storage capabilities and their limitations. We briefly evaluate the concept of object-drives and its applicability as an OSD to reduce the adverse impact for object management. Based on our analysis in Section 5, we believe that leaner stack solutions such as Object-Drives can lead towards 20-38% end-to-end time-savings over traditional OS backed object-stores. Object-Drives could also lead to reduction in host-resource consumption along-with increasing the in-node scalability while achieving higher device bandwidth [1], [11], [12].

## 2 BACKGROUND AND MOTIVATION

The backend storage servers of popular object-stores such as Lustre [3], Openstack Swift [13], Ceph [4], and MinIO[8] are built on top of traditional OS. Fundamentally, the object-storage software stack can be broken into **(1) User-space, (2) OS Kernel**, and, **(3) Storage** as shown in *Figure 1*(a).

**(1) User-space:** User-applications and OSMs form the *user-space* as they run as host processes on the storage node. OSMs manage distributed services such as application interfaces, load balancing, consistency, and fault-tolerance [3], [4], [6]. OSM translates user requests (Object-IO) received over network into locations/files in storage servers. It maintains data and metadata indexes for objects using key-value stores while orchestrating object-store operations. Similar to other distributed systems, they can also employ inefficient complex journaling on top of consistency provided by file-systems [4], [6].

**(2) OS Kernel:** In object-stores the OS acts as a middleware between userspace and storage. The OS is responsible for the persistence and data (and metadata) management in storage devices. It provides mechanisms for most translations, consistency, allocation and scheduling of SW and HW resources for physical data management. Briefly, the OS receives File-IO requests from OSM daemons where the VFS provides uniform interface to file-systems. The file system re-indexes the requests creating its own metadata such as inodes and journals. Further the request is translated into multiple Block-IOs containing the LBAs, which is scheduled by the block IO layer to the storage device via device drivers.

---

[1] **Fragmentation or aging** is an adverse phenomenon in which logically related blocks are allocated non-contiguous address space in a storage device. This can be inter-file, intra-file, or free-space fragmentation. It implies loss of *sequentiality*. The sub-optimal placement results into slower read operation due to disk seeks as well as further fragment data placement.

**(3) Storage:** Storage devices (here, OSD) are responsible for actual physical storage and provide retrieval mechanisms of the stored data. The storage device receives Block-IO requests (LBAs) from the OS device driver and internally translates it into LBA→Physical Page Addresses (PPAs) managed via in-device index management. In modern data centers there are a plethora of storage devices from mechanical HDDs to flash based SSDs to emerging byte-addressable NVMs. Each have their own characteristics and role in the storage topology. However, a common observation in production systems has been that their performance is largely influenced by the above SW layers which is agnostic to the storage device type [6], [14], [15]. We limit our scope to HDDs and SSDs which form the bulk of OSDs in production.

## 2.1 Impact of OS I/O Software Stack

Owing to the importance of the OS storage stack in the overall performance of the object-store, we discuss the impact of the IO layers followed by role and layer-by-layer breakdown.

### 2.1.1 Device Utilization & IO layer Correlation (DUiC):

To understand the impact of IO layers, we measure the throughput of block-based NVMe SSD. Using flexible IO tester (*fio*), we performed sequential write IO benchmark varying the request sizes and no. of concurrent threads on the raw-device (**RAW**) and with XFS file-system (**FS**) as shown in *Figure 2*(a). We compare the throughput for small (4K) and large (128K) request size with and without file-system while varying the number of concurrent threads to understand the requirements to saturate device bandwidth. From *Figure 2*(a), based on our observation we have the following take-aways.

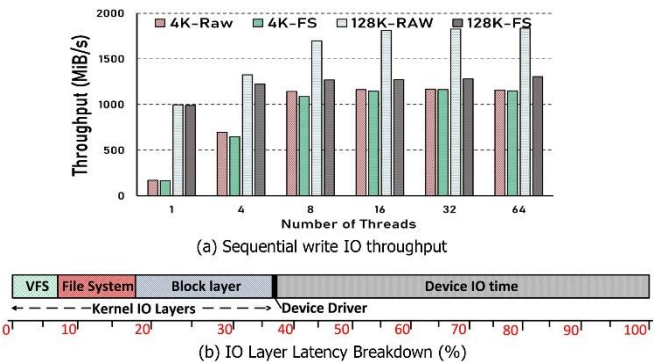

(a) Sequential write IO throughput

(b) IO Layer Latency Breakdown (%)

Figure 2 Impact of IO-layers: (a) Sequential write IO (fio) throughput on block NVMe device with (FS) and without (RAW) file-system.; (b) Normalized time-spent by an IO request in an IO-layer.

**Takeaway 1:** For the same size IO request, larger number of threads are required to achieve high throughput for file-system (more IO layers) compared to raw device. Indicating *more the layers in the IO stack, lower is device bandwidth and more host resource are consumed for increasing the device bandwidth utilization.*

**Takeaway 2:** The maximum bandwidth achieved by file-system based IO is capped and significantly lower to raw device performance. This suggests, *with increasing IO layers there are higher request processing overheads such as higher resource and lock contentions, translations, etc. which forms a bottleneck.*

**Takeaway 3:** *Larger IO requests (here 128K) are more desirable* as they can saturate device bandwidth with fewer number of threads and requests as they require lesser processing overheads such as fewer data packing queuing, etc., though the average latency increases [1], [15].

Findings in [14] also suggest that higher device bandwidth is achieved by large IOs (contiguous/sequential LBAs, e.g.128K) than small IOs (4K), for both HDDs and SSDs. Therefore, *larger sequential IOs are highly desirable for harnessing the device potential[2].*

### 2.1.2 Layer-by-Layer (LbL) Analysis:

Based on our above analysis, we understand the impact of IO layers on device utilization, we further briefly discuss first theoretically and then empirically the role of the major performance critical components of the OS Kernel IO layers.

**(1) Virtual File System:** The VFS provides the software abstraction and uniform interface to applications through system calls for accessing physical file-systems (*fs*) across storage devices. Major functionalities include concurrency control, permission checks, buffer & metadata caching [16], [17]. For performance, VFS manages 3 types of metadata caches, *i.e.,*

- The file-system (*fs*) superblock is stored in memory during mount.
- The *dentry-cache* (dcache) stores the pathname (or filename) and inode information, which aids in fast look-up to translate pathnames to dentries and validating the runtime dynamic state of files (or objects) during concurrent access.
- The *inode-cache* (icache) stores the file attributes [16].

In an object-store node, due to the large number of lookups associated to object operations typically results into large no. of small, random and sporadic metadata IOs. Therefore, *effective VFS caching plays a vital role in reducing disk I/O traffic for fs metadata operations (IO amplification as well).*

---

[2] Larger IOs increases bandwidth utilization for HDDs & SSDs [14].
 HDDs: Less disk-head seeks, lower command processing overheads→lower delay.
SSDs: May be attributed to utilization of internal flash-channel parallelism,

fewer FTL translations, lower traffic, etc [14], [35]. Though the large (sequential) & small (random) IO performance difference for SSDs is not as huge compared to HDDs.

**(2) File System (FS):** Filesystems manages the organization of data, metadata, and extended attributes (inodes, xattr). It provides POSIX-compatible common interfaces to applications for data access on storage devices. Briefly, file-system's major responsibilities include file/block allocation, metadata organization, concurrency control, File-IOs→Block-IOs translations for scheduling by block IO layer. These are achieved by employing efficient data-structures such as B-trees, log-structured, etc., for on-disk representation of data and metadata. Modern file-systems guarantee crash consistency through journaling, with advanced file-systems such as Btrfs [18] also support transactions and checksums. However, on observation the most common design principle of popular file-systems such as xfs, ext4, zfs, Btrfs [18]–[21] is to keep related data together. Thereby reducing the impact of **fragmentation or aging**. Aging effects nearly all filesystems and storage device types, resulting into loss of sequentiality for data placement and retrieval of related data. Most filesystems employ multiple fragmentation mitigating strategies such as *extents*[3], *group allocations*[4], *delayed allocation*[5] (please refer [14], [22] for details). Despite such advanced optimizations due to workload characteristics filesystems are highly vulnerable to aging. This results into sub-optimal file placement, performance degradation, poor data and metadata layout which also inhibits the ability to cache or prefetch data efficiently [6], [14], [18], [22]. Thereby, *consuming higher compute, memory & disk (more IOs) resources with increased request processing and book-keeping overheads.*

**(3) Block IO Layer:** The block-IO layer is the final OS optimization layer before dispatching IO requests to the storage device via the device driver. It receives IO requests after file-to-block mapping/allocation by the file-system, which are then translated into Block-IO (BIO) request structure which is an in-kernel data-structure representing discrete IO events, i.e. linked-list of contiguous LBAs (please refer [23] for details). It employs a block IO-scheduler which is responsible for dividing the ``*request-queue*'' bandwidth amongst contending applications, where BIO requests are staged, thereby providing opportunities to be coalesced, merged or ordered to form larger IO requests and maintain sequentiality. This shapes the final workload observed at the disk interface, depending on the IO scheduling policy. Vanilla Linux comes with CFQ, noop, and Deadline scheduling policies which prove to deliver suboptimal performance in production [15], [23]–[25]. Solutions such as Blk-mq [24] and BID [15] optimizes the block-layer for IO scheduling to cater cloud environments.

To understand the contribution of an OS layer to the overall IO request latency, we conduct **LbL** analysis using kernel probing with eBPF. The normalized layer-wise IO latency breakdown for the time-spent in each OS layer for large 128K IO request with xfs file-system (64 threads) is shown in *Figure 2*(b). We observe that on average, the time spent by an IO request in Kernel IO layers is roughly **35-40%** of the total time. While the majority of the time in the kernel is spent in the block-layer and the file-system, suggesting file-block allocation, resource (lock) contentions, queue scheduling, multiplexing (interleavings) and serialization delays of BIO requests[15] to device with increased average latency.

*Takeaway 4:* Nearly 40% of the device capability is underutilized due to the OS kernel request processing delays.

Based on our discussions, *the file-system is the most critical OS layer* as it governs device performance, host resource consumption and majorly influences working of the VFS and the block layer. In *Filesystems*, the block allocation, metadata (inode) management, fragmentation mitigation, concurrency control mechanisms influence the caching workload of VFS, and the number, size and spatial IO workload on the block-IO layer. Therefore, *we focus our attention towards the impact of file-systems deployed in production object-stores.*

## 2.2 Object-store and Choice of File-systems

Object-store user-applications and OSM workloads employ highly taxing operations (*ops*) which have tightly-coupled consistency and performance requirements.The major functionalities include from basic IO, read-modify-write, transactions, cloning, to critical reliability ops such as enumeration, fast point queries, recovery, and scrubbing, etc. Object-stores use key-value stores to build indexes. All these operations are extensively data and metadata heavy (object and local filesystem, both) [1], [6]. Apart from object persistence/access there would still be a large number of small metadata IOs.

Object-stores such as Openstack Swift [13] use filesystems extended attributes (xattr) for storing object-metadata and data as binary objects. To specifically manage object-metadata, metastores are built on over DBs such as LevelDB, RocksDB[26], etc., which consume huge host resources and limit scalability [1], [6]. Therefore, in object-stores the role of file-systems is highly prominent which in a directory are placed in same group till possible. Allocation groups (AGs) or bands concept realized by xfs (default: 4) allows superior concurrency support with the intention of minimal interference with each other.

---

[3] *Extents* are physically contiguous blocks allocated by file systems, which try to increase locality, reduce metadata and book-keeping overheads. Extents allocation help in maintaining sequentiality for larger files, as well as for small files bin-packing heuristics such as packing small files and metadata together into fewer blocks or extents, etc can lead to preserving locality.

[4] *Allocation groups (AGs)* are best-effort approaches to maintain directory locality or co-locating files in same directory till allocation group space is exhausted. Each allocation group consists of data-structures about its inodes and bitmap of blocks, with every new directory placed in an allocation group with higher number of free inodes, while inodes and data

[5] *Delayed allocation* are data placement strategies in which data blocks are buffered in-memory to batch writes and updates, and allocation to blocks occur on flushing till the consistency and durability requirements of file system or applications (fsync or flush) are not violated. Thereby increasing chances for contiguous allocation, reduction of amplification, coalescing requests together

necessitates reliability, performant data access with efficient index management. Therefore, *an object-store is as good as the local file system*.

Similar to other distributed file systems, most popular object-stores such as Ceph, Openstack Swift, MinIO support and recommend **ext4** and **xfs** file-systems which are most widely used due to their deployment popularity and proven persistence. Some salient features are as follows:

- **ext4** is de-facto file-system in most linux distributions. ext4 is update in-place filesystem which manages data in 128 MiB *extents* (runs of contiguous space) - reduces metadata book-keeping overheads. Uses tree-based index structures to represent files, and, directories for efficient block allocation tracking. ext4 uses a write-ahead journal to ensure atomicity [18], [20].

- **xfs** is also update in-place filesystem designed to support high scalability, concurrency, and parallelism. The inodes and free space information are managed by respective B+ trees, where inodes keep track of their own allocated extents [19] with the goal of reducing the amount of metadata.

- Both, ext4 and xfs employ anti-aging strategies such as extents and delayed-allocation, while **xfs** also tries to place sub-directories in different allocation groups (or bands) for concurrency and future expansion of files. Analysis of both have shown significantly lower IO amplification cost for data and metadata operations.

Owing to the semantic gap between between Object-IO and Block-IO, object storage systems employ expensive consistency mechanisms such as maintaining write-ahead-logs (WAL) on top file-system metadata journals. This increases IO amplification and synchronization dependencies between IOs resulting into high host resource consumption (and contentions). Further, it has a crippling effect on the over-all storage subsystem by fragmenting the OSD, ineffective caching, and increased page-faults for the lifetime of the OSD. *Thereby*, request processing delays can cause stallness or slowdown of computation waiting for the data [1], [6], [11], [15].

Advanced file-systems such as Btrfs [18] exposes internal transaction mechanisms to applications, while providing deduplication, checksum, and compression support which are not exposed by native filesystems such as **ext4** and **xfs**. These functionalities of Btrfs's would highly benefit object-stores by providing fast and efficient transactions with consistency guarantees. However from practical experience of Ceph team [6], Btrfs suffers from severe data and metadata fragmentation, costly checksums with lack of rollback transaction mechanisms and poor interfaces often plagued by data loss and file corruption. In the next section through empirical analysis we understand *what, why, & how* the architectural decisions at local in-node OS influences the over-all lifetime performance of objectstores.

## 3 UNDER THE HOOD ANALYSIS

Through cloud application and remote cloud storage literature [5], [7], [9], [10], [27], [28], the workloads experienced by end object storage nodes is highly skewed with multiple concurrent application submitting requests at the same time in conjunction to OSM workloads. This leads to contentions at all SW and HW resources imposing great strain on the legacy components of the traditional IO stack with repercussions on the performance and health of the object-storage system for its lifetime. In this section, through comprehensive characterization of an in-node OS object-storage stack we establish the relationship between *what and how* in the OS kernel leads to the underutilization of the OSDs deployed. Whereas the *why* was discussed in the previous sections. We broadly classify workloads into two categories:
1) Object-store core-functionality (OSM) workloads;
2) User-application workloads.
The former are services required for th object-store to function, while the later are workloads run on top of object-stores by user-applications.

First we describe the evaluation environment followed by empirical **Layer-by-Layer (LbL) analysis** as discussed below.

### 3.1 Evaluation Environment Set-up

We evaluate the in-node OS storage backend of our distributed MinIO[6] object-storage cluster. The cluster comprises of 4x storage servers with Ubuntu 16.04 LTS (xenial) OS using kernel 4.4.0-142-generic. Each storage node consists of 2x Xeon E5 2.30GHz with 48 CPUs, 128GB DRAM, 4x 15K SAS HDDs and 4x block-based SSDs. The erasure code is set to Reed-Solomon (12,4) based on cloud literature [29]. To emulate user-applications 3 clients compute nodes are connected to the cluster via 1Gbps network. For analyzing system level performance, we collect system event metrics and traces across the different layers of the IO stack for the node and OSDs using kernel spoofing tools such as BCC eBPF iovisor [30], ftrace, blktrace, strace, sar, and iostat, etc. while running representative workloads for both ext4 and xfs file systems. It is important to note, that our goal is not to compare different file-systems but to evaluate compounding impact of different architectural design decisions (*pros and cons*) which leads to underutilization of local OSDs. This also serves as a guiding stepping stone for critical design considerations for Object-Drive based object-stores.

---

[6] MinIO [8] is a popular open-source high-performance s3-compatible object-store, written in GO and assembly language with features such as inline erasure code, bitrot protection, compression, encryption, encryption with strict consistency. It only supports xfs and ext4 file system mounted OSDs.

## 3.2 Evaluation Metrics

From the discussions in Section 2, we believe the following metrics can be used to quantify and characterize the impact of the OS layer in shaping IO workload on OSDs.

- Overall performance by the workload's *total execution time*.
- VFS by metadata caching through *dcache hits/misses*.
- Filesystem by no. of page-cache evictions & no. of disk IOs for index management; *and* block-IO size distribution, size of sequential LBA chains in submission order and LBA disk-layout for block-allocation strategy.
- Block IO layer coalesces, packs and forms BIO structures by sequential LBA chains and block-IO submission request size.

For sequential chain analysis, i.e. runs of request with adjacent (numerically consecutive) LBAs for an object is considered. For e.g.: During object writes (allocation), if object size (O) is 128MB on an objectstore with 16 disks ($D_i$) set-up with Erasure-code **RS(12,4)**, i.e. 12 data (D) + 4 parity (P). There are 16 chunks of an object (O) with each chunk (C) is stored in a single OSD. In an ideal scenario, the filesystem should allocate 10MB of contiguous LBA space for object-chunk (C) inside an OSD. i.e.,

$$\frac{(O \times (D + P))}{(D \times Di)} = \frac{128\ MB \times 16}{12 \times 16} \sim 10\ MB$$

However to quantify *fragmentation or aging*, we also use another metric, **Natural Transfer Size (NTS)** proposed by Convay et.al [14], [22]. NTS corresponds to the length of sequential LBA chains submitted to the device for achieving maximal throughput performance. Any LBA (or IO) chains greater than NTS achieves full device bandwidth. Based on their analysis, NTS is 4MB for both HDDs and SSDs.

### 3.3 OSM Workload: Lexicographic Listing

Lexicographic listing or enumeration of objects is a classic object-store core functionality used extensively by OSMs and user-applications [3], [4], [6]. OSMs use it for recovery, indexing, rebalancing and scrubbing for reliability. While user-applications use it for building their own indexes. Enumeration involves access to storage and then computation (**listing + sorting**) while building indexes in cache. It is metadata heavy which involves reading object metadata (OM) or corresponding *filesystem metadata*. Each OM read involves path-lookups, sys calls, *dcache* and inode reads. This workload stresses the metadata (*inode*) management and VFS.

We write 1 million 128KB objects to the cluster with 3 clients using YCSB key-value generator [31], then enumerate all objects with cold cache with xfs and ext4 filesystem. *Figure 3* shows the metrics to evaluate the workload on an OSD during the enumeration workload. The performance of both filesystem backed objectstores take hours for completion but xfs is 3x faster than ext4 as shown in *Figure 3*. While ext4 incurs 60% lesser dcache misses than xfs (please refer *Figure 3* (b)) with denser page-cache evictions for ext4 (*Figure 3* (d)). This is attributed to the large number of page faults or expensive disk IOs (15x) for serving the same workload on ext4 compared to xfs OSD as shown in *Figure 3*(c) rendering the filesystem OS page-caching and VFS caching inefficient. On deeper inspection of LBA access pattern and heatmap during enumeration shown in *Figure 4* we visualize the superior block allocation and metadata access of xfs over ext4. We observe that xfs places metadata on specific *bands* or allocation groups while ext4 has poor metadata on-disk layout. Further, on observing the cumulative distribution (CDF) of IO size in *Figure 3*(e), most ext4 IOs are 4KB and for xfs 40% of all IO requests are 16KB. As this workload is metadata heavy, most IO requests are 4KB and 16KB in size which is also coherent to the default metadata B-tree node size for ext4 and xfs, respectively.

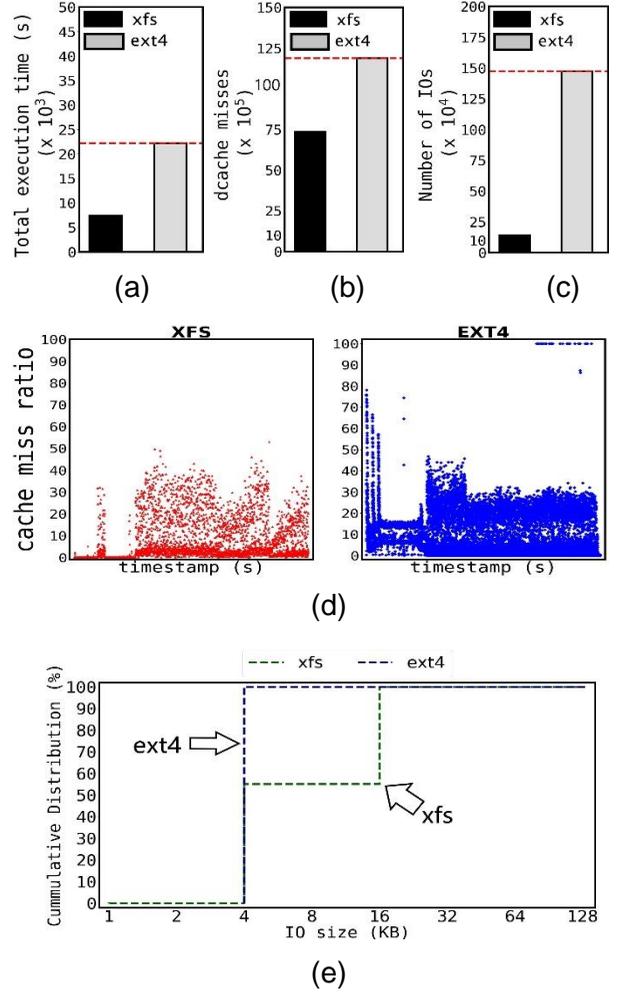

Figure 3 Object Store services Management Workload: Lexicographic Enumeration (a) total execution time; (b) dcache misses per server; (c) number of IOs incurred; (d) page cache miss ratio; (e) Cummulative Distribution (CDF) IO block-size (KB) for xfs and ext4.

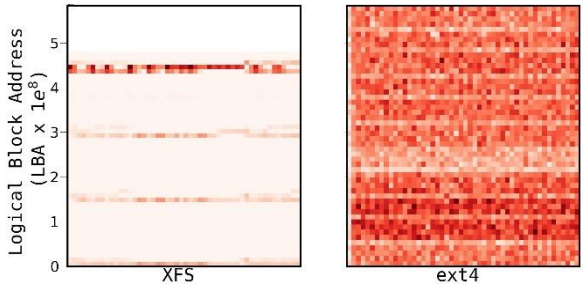

Figure 4 LBA heatmap over time during enumeration.

For xfs, this in addition to band-based allocation is the reason for better cache performance and lower page-cache misses. Therefore, this clearly shows *superior file-system metadata management results into better performance, lower resource consumption (cache, memory, disk IOs) and on-disk layout beneficial for the entire life-cycle of the object-store.*

### 3.4 Application Workload Analysis

Literature on large scale datacenters and cloud analytics services over objectstores [5], [7], [9], [10] broadly classifies workloads into three categories, namely,1) Write-Only (**W-O**), 2) Read-Only (**R-O**), and, Read-Write (**R-W**), with characteristics as shown in *Table 1*.

Table 1 Object-Store User Workload Characteristics

| Workload | Characteristics | |
|---|---|---|
| | Ops[1] | Application (object size: 128MB) |
| W-O | 95:5 | Data ingestion,enterprise backup |
| R-O | 5:95 | OLAP DW, video sharing |
| RW | 50:50 | ETL, shuffle, stage-read/writes |

[1]Object operation (REST-API) ratio:- PUT (P) : GET (G).

For simulating near realistic cloud representative workloads, we use YCSB[31] MinIO workload engine for generating object key-values with 3 clients and 128 threads concurrently issuing Object-IO requests over s3a interface to our distributed MinIO object-store cluster[7]. The size of each object (O) is selected to 128MB, which is common object size for data-warehousing and analytics objectstore workloads such as HDFS over S3 (HDFS most popular block size = 128MB). W-O is a typical bulk data ingestion workload with uniform distribution to generate similar patten as observed in [10]. Similarly, for R-O and R-W, YCSB-B and YCSB-A with zipfian distribution to emulate highly concurrent read heavy data-warehousing and extract-transform-load (ETL) workloads, respectively. These workloads are designed to simulate similar workloads on in-node OS and OSDs as experienced in highly complex and parallel cloud deployments.

**W-O workload:** W-O workload persists objects ensuring proper organization of data for efficient storage and retrieval. An object write follows *read-modify-write* protocol, *i.e*, first the object metadata (OM) is read from caches or OSDs for validity of the object and then store subsequent object data(+metadata). For an OSD, the object chunk (C) for erasure-coded systems is stored. Each require validation checks of the associated file in the OS filesystem such as path-lookups, stat, read/write system calls, etc. The IO path is redundant and depends on reading metadata and then start issuing write IOs for object data.

*Figure 5* [a-e] quantifies data placement in a write intensive workload. Both xfs and ext4 objectstore backends have marginal (~5%) difference in execution time (please refer *Figure 5*(a)), attributed to the synchronous dependency along with the OS kernel overheads. The filesystem metadata grows enormously due to large no. of files created for the objects. As shown in *Figure 5*(b), the VFS metadata caching is not effective in both cases (xfs is better) resulting in large number of disk seeks for corresponding metadata depending on filesystem index management [16], coherent with the page-fault density in *Figure 5*(c) . On analyzing block IO size CDF (*Figure 5*(d)) submitted to the OSD, predominantly all read IOs are metadata reads and 40% of write IOs (≤16K) are updates to object metadata and filesystem, with xfs (B-tree metadata node size = 16K) incurring 2x lesser metadata IOs than ext4 (node size = 4K). This is coherent with lower page-cache and dcache misses in xfs due to larger and fewer metadata.

Filesystems typically rely on OS pagecache for buffering writes and prefetching. In a lightly loaded system, writes are staged in-memory and provide the opportunity to allocate contiguous LBAs before synchronizing at regular intervals. On analysis of the write IO block size for both cases in *Figure 5*(d), we observe that roughly 50% of the write IO traffic is 256K (Object-IO) with 65% for xfs, 256K is the default max. BIO size. The 10% of the IOs are between 64K-256K indicating contentions and breakage of 256K requests into smaller IOs. This is attributed to sporadic cache evictions (write-back) which are dependent on various complex set of policies, which also lead to sub-optimal block allocation, and an object can be fragmented (non-contiguous LBAs).

On investigating the filesystem's and block IO layer's ability to retain Object chunk (C=10MB) in an OSD or submission order, we analyze the sequential chain length in

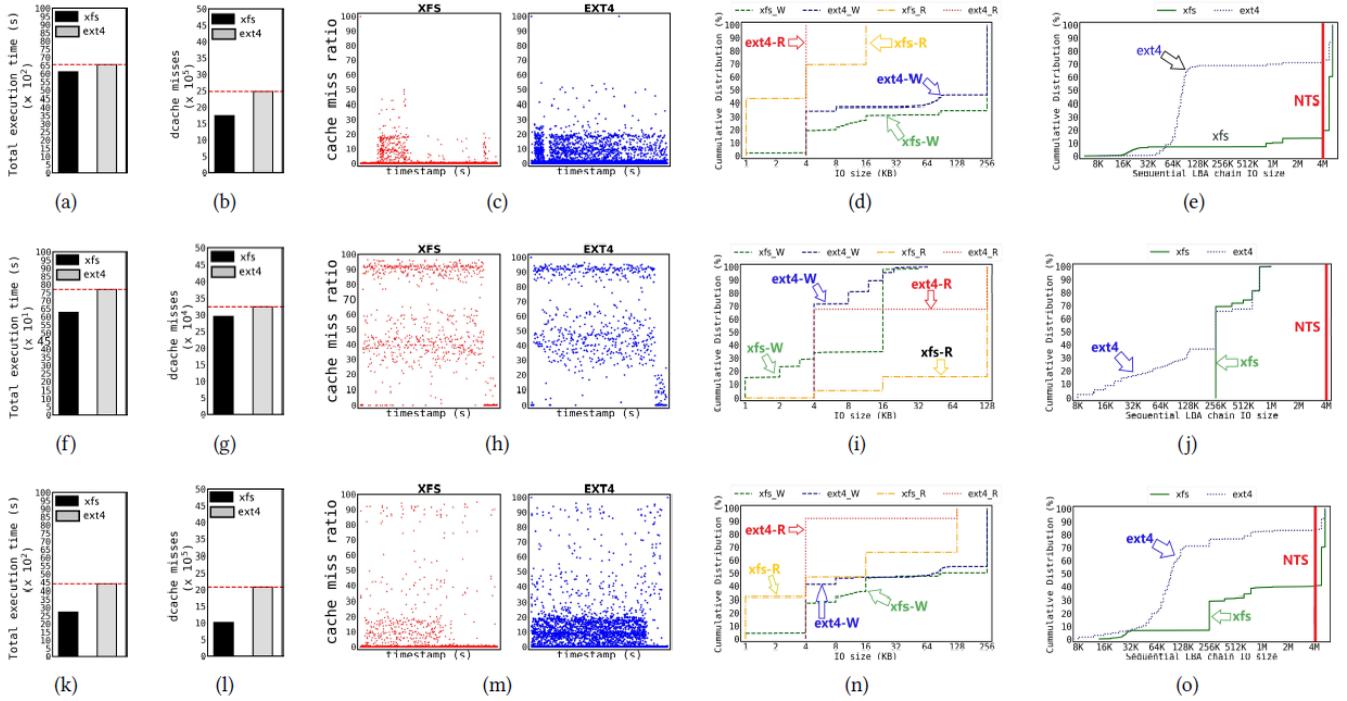

Figure 5 Application Object-storage Workload: [(a) - (e)] W-only; [(f) - (j)] R-only; [(k) - (o)] R-W as per Table 1 with ext4 and xfs backends on a MinIO distributed object-storage node. Here (a),(f),(k) total execution time; (b),(g),(l) no. of dcache misses; (c),(h),(m) page cache miss ratio; (d),(i),(n) Block-IO size CDF Write IOs:xfs,ext4, Read IOs:xfs,ext4; (e),(j),(o) Sequential chain length CDF (KB): xfs,ext4; OSD Object chunk (C) = 10MB; NTS = 4MB [Convay et. al].

*Figure 5*(e). It clearly shows that the chains are both filesystems are not able to sustain the contiguous chains and the submission order is fragmented with atleast 3 contiguous LBA fragments. For quantifying it with **NTS**, xfs has 80% of the chains greater than 4MB (NTS) while ext4 sustains only 30% with most of the chains (~70%) being less than 128K indicating poor concurrency control and higher resource contentions at block layer. Therefore, *longer IO path, dependent IOs, poor metadata management leads to sub-optimal placement, fragmentation and higher resource consumption*. The impact of fragmentation or poor data layout is more prevalent during reads.

**R-O workload:** Serving reads faster is a crucial issue in computing systems as applications need to wait for the data. Read-only workloads scan the objects from the OSDs to the compute nodes. The basic principle is to efficiently place the data in OSD such that retrieval performance is maximized. *Figure 5* [f-j] shows the characterization of R-O workload on already persisted objects. The overall execution time of xfs is 20% lower than ext4, coherent to superior metadata management and data placement being extremely vital for future retrieval during object-store lifetime. OS page-caching and VFS caching (*Figure 5* (f) and (g), respectively) are ineffective, which may be due to translations from Object IO → Block IO. However it is still 10x lower than W-O workload as an object read involves finding the relevant object file and reading it from OSD directly. CDF of block IO size in *Figure 5*(i) shows that xfs has much lower reads compared to ext4, which constitutes 20% (≤16K) and 80% (≤4K) of the IO traffic respectively, rest is object data (128K). Sequential chain analysis in *Figure 5*(j), shows that object sequentiality is not retained, with xfs having superior index organization and concurrency control with 70% of chains being 256K in size compared to 40% for ext4. In best case, 20% of the chains are 1MB, much lower than NTS. The intense contention at block layer makes coalescing ineffective with max. 8 IOs for xfs (1MB/128K) and 2 IOs for ext4 (256K/128K) being merged. This clearly indicates *read-aging* and underutilization of the OSD.

**R-W workload:** In R-W workloads (similar to ETL), objects are read, computed and, if required new objects are persisted. *Figure 5* [k-o] characterizes the R-W workload, we observe that the execution time xfs is 1.4x faster than ext4 with 2x lower dcache misses. While page-cache is ineffective with spikes of nearly 100% misses due to reads and density of ext4 misses much higher than xfs. This is coherent to the CDF block IO size findings (*Figure 5*(n)) with 90% and 50% of read IO traffic for ext4 and xfs respectively, is due to metadata lookups. On analyzing sequential chains in *Figure 5*(o), even in best case i.e. xfs 40% of the chains are shorter than NTS (4MB) while during data-placement (W-O) its 80% (for ext4 its 15%). Therefore, similar to R-O workloads, R-W are also impacted by initial data layout due to *read-aging* during reads for already persisted data and suggesting *free-space fragmentation* due to unavailability of runs of free space for new objects due to already poorly organized data in OSDs.

Through our comprehensive workload characterization it is clearly evident that complex and redundant object management policies coupled with application requirements puts intense strain on the already vulnerable OS stack. This leads to the underutilization of the OSD (object-storage device) deployed for the lifetime of the objectstore. Though, both filesystems in consideration (ext4 and xfs) suffer from *read-aging*, however due to superior metadata organization and data allocation strategies xfs outperforms ext4 while consuming lower host and disk resources. Therefore, **a conventional OS based object-store is as good as the underlying OS IO stack**.

## 4 RELATED WORKS

There have been a plethora of solutions proposed for reducing the impact of long and iterative OS IO path. From developing new file-systems such as BtrFS [18], BetrFS [32], F2FS [21] for efficient index and consistency management to storage device specific optimizations [32]–[35]. To reduce index management overheads, multiple solutions are proposed from use of hash-based operatives such as DLFS [36] to bypassing some OS layers such as ByVFS [16] which tend to avoid VFS for metadata overheads. File systems have been discussed in detail in the paper. While advanced optimizations such as atomicity, transaction, and, snapshot support to local file systems as well as use of userspace filesystems such as FUSE [37], FSaP [17] have been futile due to unstability in production environments [6]. The adoption of data-centric specific emerging devices requires extensive revamping of the host and data management storage stacks to overcome OS data path limitations. For instance, recent interests in the development of device specific optimizations such as [38] for high capacity SMR HDDs, and Open-channel SSDs and Zoned Namespace SSD (ZNS) [39] to eliminate long FTL IO tail latency. Local production file systems such as ext4, and xfs work on update-in-place design, and the zone management requires log-structured COW, making the interfaces backward incompatible. While their host-managed solutions [38] is highly unstable with unpredictable performance with control plane shifting towards host resulting in higher resource consumption [6]. There have been also been efforts to build complete storage subsystems without kernel filesystem and caches such as Aerospike [40]. Initial Ceph [4] object-store- FileStore is built on top of file system, preferably xfs. Due to the afore-mentioned issues with redundant OS IO stacks lead to the development of Ceph BlueStore [6]. BlueStore is designed to manage indexes or object metadata using ordered key-value store RocksDB [26] and runs of custom built userspace filesystem BlueFS, while the binary object data is written asynchronous to raw block devices for maximizing *sequential* allocations. KV stores such as RocksDB can improve metadata performance but consume huge amount of host resources limiting the innode scalability, suffer from severe compaction and high write-amplification limiting the lifetime of the OSD [1]. All these solutions are highly specific limiting their adaptability in production storage systems, most of them being piece-meal solutions which have high tradeoffs between host resource consumption, stability, consistency and high performance [1], [6].

## 5 FUTURE DIRECTIONS: OBJECT DRIVES

Considering the complexity of building storage systems from scratch, a natural progression is the development of holistic storage device technologies which can execute host data management functionalities. In production eliminating the OS kernel completely is not feasible but offloading some key performance critical functionalities with minimal infrastructural changes can provide seamless benefits (please refer Section 4). Here, we briefly discuss the potential of such a new class of storage devices, defined as Object-Drives as a solution to alleviate object-stores from afore-mentioned hefty and redundant host (OS + application) IO-path discussed in the previous sections. In the case of object-drives, host data management functionalities can be offloaded and execute inside the storage device itself, thereby simplifying the overall SW storage stack and streamlining the data processing.

Samsung KV-SSD [1] and Seagate Kinetic Drive [2] are classic industrial use-case implementations of Object-Drives which use a key-value (KV) interface instead of traditional block based (LBA) interface. Seagate Kinetic platform [2] provides a KV interface over Ethernet while executing LevelDB KV-store inside a HDD. Samsung KV-SSD [1] is the first Key-Value SSD (KVSSD) industrial prototype with SNIA standardized API, which executes key-value management inside the device itself. Host kv-stores can directly interact with underlying devices using KV-pairs through a thin host library KV-API, thereby allowing users to switch from block SSDs to KVSSD with just a firmware change.

Key-value stores such as RocksDB [26] are extensively employed by storage backends, especially objectstores for basic object (data (OD) + metadata (OM)) load/store, caches, consistency and indexing. We briefly discuss the potential of Object-Drives deployed in objectstores as OSDs. We limit our discussion to alleviation of OS overheads while being storage media agnostic as we want to evaluate the concept of Object-Drive. However, we use KVSSD [1] as an use-case which has recently gained attention in industrial and academic research [1], [11], [12], [41]. Consider *Figure 1*(b) which shows our proposed Object-Drive based object-storage architecture that is also broadly divided into three major components:

- **Userspace** consists of user applications and objectstore manager (OSM) services for managing user object-namespace.
- **OS kernel** consists of light weighted Object-Drive library which acts as a glue between userspace and the OSD, providing an interface to persist Object KV pairs similar to KV-API in KVSSD [1]. Therfore, shifting the data control plane to the OSD to be managed internally.
- Object-Drive as OSD, which can support variable

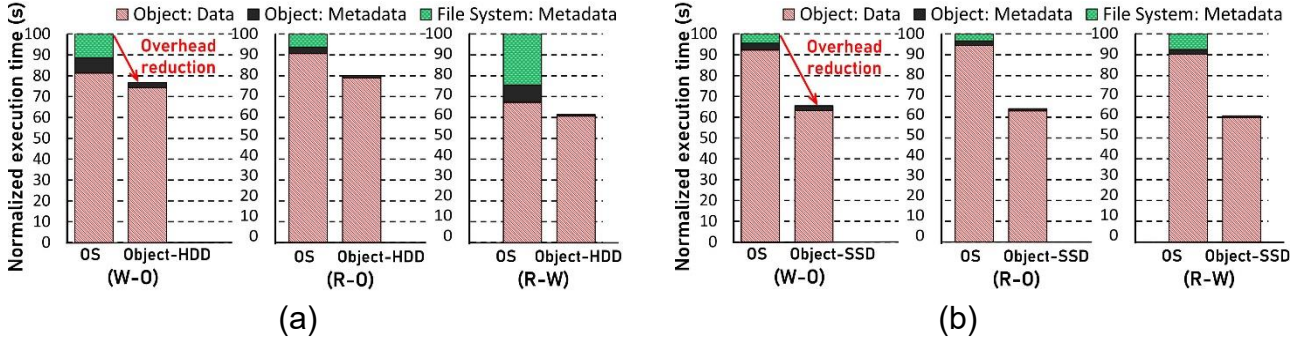

Figure 6 Normalized latency breakdown for user-applications comparing conventional OS-based with Object-Drive OSD with (a) HDD; and (b) SSD, storage media backend. IO type: Object Data (OD), **Object Metadata (OM)** and Filesystem Metadata (FSM).

sized KV-pairs internally managing Object-IO → Physical Page addressing via multi-level hash via flat-namespace such as defined in DLFS [36]. The evident advantage of using Object-Drives is that it makes the object-storage stack much leaner compared to traditional OS based stack shown in *Figure 1*.

Object-Drive solutions are capable of alleviating object-stores from the adverse issues due to the hefty operating system (OS) IO stack, as discussed in Sections 2 and 3 due to the following:

*1. Light weight address translation:* Expensive multi-layer translation mapping is not required any more. Application characteristics can be retained with variable size KV pair support contrary to fixed sized (LBA) blocks in block devices, The IO path is reduced to Object-IO → Object-ID (OID) to identify KV-pairs by the device. Clearly, host resources of compute, memory (VFS, caches) and expensive disk IOs are saved (discussed in Section 3) for data and metadata. OIDs as key can directly be used by applications via the Object library and type of request in the command header.

*2. Simplified persistence:* Data in our use-case is large to increase the chances of placing related data together (or reduce *aging*) inside the device, the driver can keep track of object-parts initiating multiple fixed size DMAs larger than device Natural transfer size (NTS) asynchronously for performance with hints from driver. Allocation of large data similar to DLFS [36] using hashes for supporting large sequential allocation with efficient bin-packing algorithms.

*3. Consistency without journaling:* Similar to KV-SSD [1] battery-backed DRAM and in-device transaction support [41] provides consistency gaurantees and in-device index managers eliminate the need for additional logging and journaling mechanisms while reducing read/write IO amplification.

*4. Advanced functionalities:* Similar to KV-SSDs [1], advanced feature-set such as in-device support for grouping keys (objects) or *iterators* can be extremely useful for OSM core-functionalities such as *lexicographic listing* discussed in Section 3.3. This reduces host resources and time for listing related objects while also frees up device bandwidth due to multiple IOs for every object.

*5. Reduced contentions:* As discussed previously, with fewer layers and simplified device internal object management, the contention for resources is significantly less. KV-SiPC [11] analyzes time spent on locks is ~35% lower for KV-SSD stacks than traditional OS IO stacks. This is due to the reduction in number and synchronous dependence of metadata ops such as on-demand indoes, caches, translations, WAL and journals compared to block-devices. Such overheads can also account for upto 70-90% of the IO traffic (Section 3.4 and [42]). Therefore, coherent to our analysis in the previous sections, i.e., *leaner the IO path lesser the overheads in request processing.*

### 5.1 Object-Drive Analysis

Based on our analysis we understand that on average atleast ~40% of the total IO time is consumed by kernel resources (please refer *Figure 2*(b)), while in complex objectstore environments it could be much higher. We focus on the fundamental root issue, i.e. need for the shift from hefty and redundant OS IO stack in an objectstorage node. As per our knowledge other works on such devices such as [1], [11], [12] focus on individual device performance and application. Therefore, *our goal is to briefly evaluate the concept of Object-Drive agnostic of the storage media type (i.e. HDDs and SSDs) in reducing the OS overheads when deployed as an OSD.*

It is outside the scope of this paper to implement an objectstore on top of Object-Drives which requires complete control and IO path re-implementations with consistency gaurantees, leaving aside the most important factor of limited public availability of Object-Drives. Hence, we conduct trace-driven simulation by collecting block IO traces, system metrics from an OSD and analysis framework using the tool-chains described in Section 3.1 during actual application workloads runs ( please refer *Table 1*). The parameters for modeling storage device characteristics is cho-sen from the data-sheet specifications on which these devices have been implemented, i.e. Seagate Kinetic HDD model ST4000NK001 [43] for HDDs and Samsung PM983 NVMe SSD [44] (KVSSD [1]) for SSDs, respectively. Based on the eBPF tool chains [30] and our analysis framework as discussed in Section 3, we identify the source and type of IO, i.e. Object Data (OD), Metadata (OM) and Filesystem Metadata (FSM) by tags, respectively. *Figure 6* shows the normalized total layer-wise latency breakdown incurred during a workload for access to OD, OM and FSM in an OSD deployed in conventional OS (xfs) and Object-Drive objectstore with HDD and SSD as OSD, *Figure 6*(a) and (b) respectively. As mentioned earlier, our goal is not to compare HDDs or SSDs but evaluate the benefits of reducing OS overheads by Object-Drives for a particular workload.

Comparing to its objectstore counterparts, for all workloads the total overhead reduction using ObjectDrive OSD is 20-38% in total time-savings. The gains are higher in write or update heavy workloads, particularly for **R-W** workload as its metadata heavy and the major benefits comes from reduction of FSM IO traffic. These are small dependent IOs (*read-modify-write*) which increases the request processing overheads as well as introduces randomness in IO affecting all layers of the IO hierarchy as well as HDDs and SSDs both. The impact on HDDs is higher as compared to SSDs (please refer the green bar in *Figure 6*(a) and (b) W or R-W workload) due to mechanical disk seeks. Therefore, direct Object access without multi-layer translations can significantly reduce delays due to FSM. The streamlined light-weighted object translations also reduces the object metadata (OM) book-keeping overheads in all workloads which results in decrease in OM IO traffic. We also observe that for servicing the actual object data (OD) in ObjectDrive solutions the time taken is relatively low. This is attributed to the in-device OID translations, decreased metadata dependencies, suggesting larger sequential chains or larger IOs (more opportunities to coalesce) during storage and retrieval in OSDs due to also removal of randomness causing FSM IOs with reduced contentions.

However, our model is not fully accurate as we do not account for latencies for key handling and packing inside the device, however at high queue depths and large no. of threads the effect is amortized [42]. We purposely do not model complete contiguous object placement inside the device as our analysis is based on OS overhead removal as actual existing devices have limitations for value size (~2MB for KVSSDs [1]). Studies such as [1], [42] show that individual IOs for KVSSDs can be lower than block the end-to-end (E2E) performance benefits due to overheads reduction is much higher with reduction in CPU utilization, i.e. ~13x compared to RocksDB on block-SSD. Our study quantifies the impact of the OS overheads in object-stores, more specifically *what, why, and, how* leads to the underutilization of the OSDs.

## 6 CONCLUSION

In this paper, through comprehensive under-the-hood analysis of an object-storage node, we characterize the impact of object-store workloads on the Operating Systems IO stack in a storage server and its subsequent rippling effect on the underlying Object-Storage Device (OSD). We observe that the legacy architecture of the OS storage stack coupled with complex data management policies leads to a performance mismatch between what an end-storage device is capable of delivering and what it actually delivers in a production environment. Owing to the associated issues of object-management and the vulnerabilities of existing IO software stacks, we discuss the potential of a new class of storage devices, known as **Object-Drives**, where host data management functionalities can be offloaded to the storage device. Thereby, making the IO stack leaner. Based on our analysis, we believe object-drives can alleviate object-stores from highly taxing overheads of data management with 20-38% time-savings over traditional OS storage stack.

## 8 LEGAL INFORMATION

THIS DOCUMENTATION IS PROVIDED FOR REFERENCE PURPOSES ONLY, AND ALL INFORMATION DISCUSSED HEREIN IS PROVIDED ON AN "AS-IS" BASIS, WITHOUT WARRANTIES OF ANY KIND. SAMSUNG SEMICONDUCTOR, INC. ($"SAMSUNG"$) ASSUMES NO LIABILITY WHATSOEVER, INCLUDING WITHOUT LIMITATION CONSEQUENTIAL OR INCIDENTAL DAMAGES, AND SAMSUNG DISCLAIMS ANY EXPRESS OR IMPLIED WARRANTY, ARISING OUT OF OR RELATED TO THE USE OF THIS DOCUMENT, INCLUDING BUT NOT LIMITED TO, LIABILITY OR WARRANTIES RELATED TO FITNESS FOR A PARTICULAR PURPOSE, MERCHANTABILITY, OR INFRINGEMENT OF ANY PATENT, COPYRIGHT, OR OTHER INTELLECTUAL PROPERTY RIGHT. SAMSUNG RESERVES THE RIGHT TO CHANGE THE INFORMATION, DOCUMENTATION, AND SPECIFICATIONS WITHOUT NOTICE. THIS INCLUDES MAKING CHANGES TO THIS DOCUMENTATION AT ANY TIME WITHOUT PRIOR NOTICE. SAMSUNG ASSUMES NO RESPONSIBILITY FOR POSSIBLE ERRORS OR OMISSIONS, OR FOR ANY CONSEQUENCES FROM THE USE OF THE DOCUMENTATION CONTAINED HEREIN.